\documentclass{article}
\renewcommand{\theequation}{\arabic{equation}}

\usepackage[dvips]{graphicx}
\usepackage{psfrag}
\usepackage{subfigure}
 
\begin{document}

\renewcommand{\theequation}{\arabic{equation}}

\begin{center}
{\Large {\bf Phase Structure of the Random-Plaquette
$Z_2$ Gauge Model: Accuracy Threshold for a Toric Quantum Memory }}

\vskip 1cm

{\Large Takuya Ohno\footnote{e-mail address:
taquya@phys.kyy.nitech.ac.jp}}

{Department of Electrical and Computer Engineering, 
Graduate School of Engineering,
Nagoya Institute of
Technology, Nagoya 466-8555 Japan}

{\Large Gaku Arakawa\footnote{e-mail address:
e101608@phys.kyy.nitech.ac.jp}}

{Department of Applied Physics,  Graduate School of Engineering,
Nagoya Institute of
Technology, Nagoya 466-8555 Japan}

{\Large  Ikuo Ichinose\footnote{e-mail 
 address: ikuo@nitech.ac.jp}}  

{Department of Applied Physics, Graduate School of Engineering,
Nagoya Institute of Technology,
Nagoya, 466-8555 Japan}

{\Large Tetsuo Matsui\footnote{e-mail address:
matsui@phys.kindai.ac.jp}}

{Department of Physics, Kinki University, Higashi-Osaka,
577-8502 Japan}
\end{center}
\setcounter{footnote}{0}
\begin{center} 
\begin{bf}
Abstract
\end{bf}
\end{center}
We study the phase structure of the 
random-plaquette $Z_2$ lattice gauge model in three dimensions.
In this model, the``gauge coupling" for each plaquette is a
quenched random variable that takes the value $\beta$ with the 
probability $1-p$ and  $-\beta$ with the probability $p$.
This model is relevant for the recently proposed 
quantum memory of toric code.
The parameter $p$ is the concentration of the plaquettes 
with  ``wrong-sign" couplings $-\beta$,
and interpreted as the  error probability per qubit 
in quantum code.
In the gauge system with $p=0$, i.e., with the uniform 
gauge couplings $\beta$, it is known that there exists a
second-order phase transition at a certain critical 
``temperature", $T(\equiv \beta^{-1}) = T_c =1.31$, 
which separates an ordered(Higgs) phase at $T < T_c$
and a disordered(confinement) phase at $T>T_c$.
As $p$ increases, 
the critical temperature $T_c(p)$ decreases. 
In the $p-T$ plane, the curve $T_c(p)$ intersects with the Nishimori line
$T_{\rm N}(p)$ at the certain point $(p_c, T_{\rm N}(p_c))$.
The value $p_c$ is just the accuracy threshold for 
a fault-tolerant quantum memory and associated
quantum computations.
By the Monte-Carlo simulations, we calculate the specific heat and the
expectation values of the Wilson loop to obtain 
the phase-transition line $T_c(p)$ numerically.
The accuracy threshold is estimated as $p_c \simeq 0.033$.

\section{Introduction}
Gauge theory plays a very important role in modern physics.
All the interactions between elementary particles are described
by gauge theories.
Furthermore, gauge theories present important and
clear understandings for other areas of physics including 
superconductivity, fractional quantum Hall effect, random spin systems,
localization by random vector potentials.

Recent developements of nanotechnology elevate the old interesting 
theoretical ideas
to realistic objects. Quantum computer(QC) is one of them.
One of the most difficult problem for constructing QC is how to
keep quantum states stable against noises 
while quantum computations are performed,
i.e., the problem of decoherence.
In order to make a quantum computation fault-tolerant, 
the involved qauntum states should have certain ``quantum numbers",
and these quantum numbers are required to stay as  good ones 
during the computation.
From this point of view, one may expect that gauge-theory mechanisms like
Aharonov-Bohm(AB) effect play an important role in quantum computation.

In fact, Kitaev\cite{kitaev} recently proposed
a fault-tolerant quantum memory and quantum computations that are 
based on the AB effect of discrete $Z_2$ gauge symmtery.
After that many interesting studies on Kitaev's model 
appeared\cite{preskill,dennis,wang,mochon,benoit,AI,takeda}.
Among them, two of the present authors\cite{AI} generalized Kitaev's model 
and also clarified the relationship between Kitaev's model and a 
{\em spontaneously broken} $U(1)$ gauge theory.

For quantum computations, it is important to estimate its accuracy threshold.
As discussed in Ref.\cite{wang},  it is
obtained for Kitaev's model by studying the random-plaquette 
gauge model(RPGM) on a three-dimensional ($3D$)
lattice with $Z_2$ gauge symmetry.
In the RPGM, the (inverse) gauge coupling for each plaquette
takes the values $\pm\beta$ with
random sign; the probability to take $\beta$ is given by $1-p$ 
and the probability of  ``wrong-sign" $-\beta$ is $p$.
The original nonrandom ($p=0$) $3D$ $Z_2$ lattice gauge theory 
is known to be dual to the classical $3D$ Ising model.
The fluctuations of gauge field are controlled
by the ``temperature" (the coupling constant)
$T \equiv \beta^{-1}$. The system 
exhibits a second-order phase transition at 
the critical temperature $T = T_c = 1.31$. 
For $T < T_c$ the gauge-field fluctuations are small
and the system is in the ordered Higgs phase, whille 
for $T_c < T$, the fluctuations are large
and the system is in the disordered confinement phase\cite{kogut}.
As $p$ increases from $p=0$, 
the critical temperature $T_c(p)$ 
decreases due to the randomness in gauge couplings, 
and it shall vanish at a certain value $p_{0}$, $T_c(p_{0})=0$. 
In Fig.1, we illustrate a schematic  phase diagram in the $p-T$ plane.
The solid line expresses $T_c(p)$, which may decrease
very rapidly near $p_{0}$.
The dashed line is the Nishimori line $T(p)=T_{\rm N}(p)$\cite{nishimori} 
on which fluctuations of gauge field caused by the quantum effect 
and those caused by the random couplings are equal.
As we explain it in detail later on, 
we consider a particular method of error corrections in QC's 
that is realized along the Nishimori line. Also the system
should be  in the ordered Higgs phase in order that 
the quantum computations are fault-torelant.
Thus the intersection point of the critical line $T_{c}(p)$ 
and the Nishimori line, i.e., $T_{c}(p_c) = T_{\rm N}(p_c)$, 
locates  the accuracy threshold for the quantum computations by Kitaev's model.
One concludes that the quantum computations are fault-torelant for $p < p_c$. 

The present paper is organized as follows.
In Sec.2, we shall review Kitaev's model of a quantum toric code
and its relation to the $Z_2$ gauge model.
In Sec.3, the relationship between the accuracy threshold 
of the toric code and the random spin and gauge models 
is discussed.
Section 4 is the main part of the present paper.
We present the results of Monte-Calro simulations of the RPGM. 
We calculate the specific heat and expectation values of the
Wilson loop for various points in the $p-T$ plane.
There are (at least) two phases in the model; one is the ordered
Higgs phase and the other is the disordered confinement phase.
From these studies, we estimate the accuracy threshold 
of the toric code as $p_c \simeq 0.033$.
This value of threshold is consistent with the previous 
studies\cite{wang,takeda}.
In Ref.\cite{wang}, the $Z_2$ RPGM in $3D$ was investigated by
studying the lowest-energy configurations for various random
gauge couplings, and then
the critical concentration along the $T=0$ axis was estimated as
$p_0=0.029$.
By using specific symmetry of the model on the Nishimori line,
it is proved that 
the above value gives a lower bound for the accuracy threshold 
$p_c$\cite{nishimori}.
On the other hand in Ref.\cite{takeda}, the {\em four-dimensional} 
RPGM was studied. The value of $p$ at the intersection point
of the phase boundary and the Nishimori line
of this model gives an upper bound of $p_c$ of the RPGM in $3D$.
 By using the  self duality of the model,
it is obtained as $0.11$, which is much larger than
our result of $p_c$. 
Section 4 is devoted for conclusion.

\section{Quantum toric code}
The quantum toric code was first proposed by Kitaev\cite{kitaev},
and many interesting studies on that idea appeared after that.
It is more transparent to formulate Kitaev's model 
as a $Z_2$ lattice gauge theory\cite{AI}.
The quantum toric code is viewed as a set of qubits put on links of 
a two-dimensional$(2D)$ square lattice with the periodic
boundary condition (i.e., a torus).  
Each link of this lattice is denoted as $(x,i)$ where $x$ is the
site index and $i=1,2$ is the spatial direction index.    
The qubit residing on the link $(x,i)$ is described by a quantum
$Z_2$ variable $Z_{xi}$ and its conjugate variable $X_{xi}$.
$Z_{xi}$ is regarded as the $Z_2$ gauge field, and so $X_{xi}$
plays the role of the electric field. 
They satisfy the commutation relations like 
\begin{equation}
X_{xi}Z_{xi}=-Z_{xi}X_{xi},\;\;\;
X_{xi}Z_{yj}=Z_{yj}X_{xi}\;\;
\mbox{for} \; (x,i)\neq (y,j).
\end{equation}
In the matrix representation, $Z_{xi}=\sigma^z_{xi}$ and 
$X_{xi}=\sigma^x_{xi}$, where $\sigma^{x}$ and $\sigma^{z}$ 
are $2\times2$ Pauli spin matrices.

Let us consider the following generalized Hamiltonian,
\begin{eqnarray}
H_{\cal T}&=& H_Z+H^\psi_Z+H^\varphi_Z,  \nonumber \\
H_Z &=& -\lambda_1\sum_{\mbox{\footnotesize{link}}} X_{xi}
-\lambda_2 \sum_{\mbox{\footnotesize{plaquette}}} ZZZZ 
+ \mbox{H.c.}, \nonumber \\
H^\psi_Z &=& - \gamma \sum_{\mbox{\footnotesize{link}}} 
\psi^\dagger_{x+i} Z_{xi}\psi_x
+M\sum_{\mbox{\footnotesize{site}}} 
\psi^\dagger_x\psi_x+\mbox{H.c.}, \nonumber\\
H^\varphi_Z &=& - \gamma \sum_{\mbox{\footnotesize{link}}} 
\varphi^\dagger_{x+i} Z_{xi}\varphi_x
+M\sum_{\mbox{\footnotesize{site}}} 
\varphi^\dagger_x\varphi_x+\mbox{H.c.}, 
\label{HZ}
\end{eqnarray}
where $\lambda_1$ and $\lambda_2$ are coupling constants, 
and $\psi_{x}$ and $\varphi_{x}$ are ``relativistic" 
fermions sitting on site $x$ with the mass $M$ 
and the hopping parameter $\gamma$.
They satify the canonical anticommutation relations like
\begin{equation}
\left\{\psi_x, \psi_y^{\dagger}\right\} = \delta_{xy}, \;\;
\left\{\varphi_x, \varphi_y^{\dagger}\right\} = \delta_{xy}.
\end{equation}
The above Hamiltonian (\ref{HZ}) is invariant under $Z_2$
local gauge transformation. In fact, we have
\begin{eqnarray}
&& G_x \equiv \Big(\prod_{(y,i)\in x}X_{yi}\Big)
e^{-\pi i(\psi^\dagger_x\psi_x
-\varphi^\dagger_x\varphi_x)}, \nonumber\\
&& \left[G_x, H_{\cal T}\right] = 0,
\label{generator}
\end{eqnarray}
where $(y,i)\in x$ denotes 4 links emanating from site $x$,
and the operator $G_x$ is the generator of gauge transformations.
Respecting the gauge invariance of $H_{\cal T}$,
the phyical states $|phys\rangle$ should be gauge invariant, i.e.,
satisfy the following physical-state condition, 
\begin{equation} 
G_x|phys\rangle=|phys\rangle.
\label{Phys}
\end{equation}

The Hamiltonian of 
Kitaev's model is obtained by setting $\lambda_1=0, \lambda_2=1, \gamma=0$
and $M=2$\cite{AI}, and then the quantum space of the gauge field  and that
of the fermions are coupled only through the gauge-invariance condition.
In this case, it is easy to see that 
the ground states $|GS\rangle_K$ satisfy the following conditions,
\begin{eqnarray}
&& \psi_x^{\dagger}\psi_x|GS\rangle_K =0, \;\;\;
\varphi_x^{\dagger}\varphi_x|GS\rangle_K =0,\nonumber\\
&& A_x|GS\rangle_K=|GS\rangle_K, \;\;\; 
B_P|GS\rangle_K=|GS\rangle_K,
\label{Grand}
\end{eqnarray}
for all the sites $x$ and plaquettes $P$,
where the operators $A_x$ and $B_P$ are defined as follows,
\begin{equation}
 A_x=\prod_{(y,i)\in x}{X}_{yi}, \;\; \mbox{and} \;\; 
 B_P=\prod_{(y,i)\in P}{Z}_{yi}.
\label{ABsta}
\end{equation}
These operators are called {\em stabilizers} or {\em check operators}.
From Eq.(\ref{generator}), 
$A_x$ is the generator of
a local gauge transformation of the $Z_2$ gauge field $Z_{xi}$,
and $B_P$ measures $Z_2$-gauge flux penetrating the plaquette $P$,
i.e., $B_P=1$ for fluxless states and $B_P=-1$ for fluxful states.
Eq.(\ref{Grand}) means that
$|GS\rangle_K$ involve no fermions (pure-gauge state), gauge-invariant,
and fluxless.

As the system is defined on the torus, the ground states are not unique
but 4-fold degenerate.
They satisfy Eq.(\ref{Grand}) and are distinguished by a pair of
``nonlocal" quantum numbers $Z_a(a=1,2)$ defined as
\begin{equation}
Z_a=\prod_{C^a_Z}Z_{xi},  
\label{Za}
\end{equation} 
where $C^a_Z(a=1,2)$ are two noncontractible closed loops on the original
lattice (see Fig.2).
The ground states have the quantum numbers $(Z_1,Z_2)=(1,1),(1,-1),(-1,1)$
and $(-1,-1)$.
Quantum informations for the toric code are stored in the 
above 4-fold degenerate ground states
which work as a {\em fault-tolerant qudit}. 
Instead of $Z_a$, we can consider a set of conjugate quantum numbers 
$X_a(a=1,2)$ given by
\begin{equation}
X_a=\prod_{\tilde{C}^a_X}X_{xi},
\label{Xa}
\end{equation}
where $\tilde{C}^a_X$ are nontrivial closed loops on the {\em dual} 
lattice (see Fig.2) and $X_{xi}$'s in (\ref{Xa}) reside on the 
links of the original lattice that are crossed by $\tilde{C}^a_X$.
It is easily verified that $Z_a$ and $X_a$$(a=1,2)$ satisfy the same
commutation relations with the Pauli matrices.
Then the ground states are characterized by the quantum numbers
$(X_1,X_2)=(1,1),(1,-1),(-1,1)$ and $(-1,-1)$.

The gauge-theory aspect of Kitaev's model becomes 
clear when we consider excitations in the system.
Excitations obviously break the local condition (\ref{Grand}) 
at some specific sites and/or plaquettes.
As the stabilizers satisfy the identities like 
\begin{equation}
\prod_{\mbox{\footnotesize{all sites}}}A_x=1,\;\;
\mbox{and}\;\; 
\prod_{\mbox{\footnotesize{all pl's}}}B_P=1,
\label{AB}
\end{equation} 
these excitations should appear in  pairs.
A simple example is the fermion-pair state with two fermions at $x$ and $y$,
\begin{equation}
|F;C_{xy}\rangle =
\psi^\dagger_y\Big(\prod_{C_{xy}}Z\Big)\varphi^\dagger_x|GS\rangle_K,
\label{Fstate}
\end{equation}
where $C_{xy}$ is a certain path on the {\em original} lattice
connecting $x$ and $y$.
Another example is the vortex-pair state with two vortices at 
{\em dual} sites $x^\ast$ and $y^\ast$,
\begin{equation}
|V;\tilde{C}_{x^\ast y^\ast}\rangle  
 = \Big(\prod_{\tilde{C}_{x^\ast y^\ast}}X\Big)|GS\rangle_K,
\label{Vstate}
\end{equation}
where $\tilde{C}_{x^\ast y^\ast}$ is a certain path on the {\em dual} lattice
connecting $x^\ast$ and $y^\ast$, and $X$'s in (\ref{Vstate})
are on the links crossing $\tilde{C}_{x^\ast y^\ast}$(see Fig.3).
Wave function of the state containing
the above fermionic and vortex excitations acquires nontrivial phase factors
when one of the excitations winds around the other.
This is the AB effect of the $Z_2$ gauge theory.

As explained above, quantum informations are stored in  a set of 
the 4-fold 
degenerate groundstates, ${\cal L}=\{|GS\rangle\}$.
During a quantum computation, quantum states in quantum memory are to be 
damaged by noises which cause decoherence of the quantum states.
As a result, excited states appear, i.e., errors occur.
One can check whether errors have occured by measuring
the operators $\psi^\dagger_x\psi_x, \varphi^\dagger_x\varphi_x$,
$A_x$, and $B_P$ in the quantum states and judging
whether the conditions (\ref{Grand}) are satisfied or not.
We can correct these errors by applying suitable operators on 
the quantum states.
For example,  the state with vortex-type excitations can be corrected
by applying the operator $(\prod X)$, a product of  $X_{xi}$ along the 
path appeared in the excited state, owing to the identity $X_{xi}^2=1$.
On the other hand, for the charge-type excitations, 
the operator $\psi_y(\prod Z)\varphi_x$ does the job.
But we note that we can know only the locations of vortices 
and/or charges but not that of the error chain itself.
Therefore we should consider specific methods of the error corrections.
As a results there exists an accuracy threshold $p_c$, i.e., 
if the error probability per qubit and 
per unit computer time exceeds $p_c$, we cannot perform successful
error corrections in practical manner.
The threshold can be estimated by studying the random-bond Ising model(RBIM)
in $2D$ and/or the RPGM in $3D$, as we discuss it in the following section.

\section{Accuracy threshold and RPGM}

It is interesting and important to estimate the accuracy threshold
$p_c$ of the toric quantum code.
The value of $p_c$ should be compared with those of other quantum
memories and quantum gates which are estimated as
$10^{-6}$ to $1/300$\cite{estimate}.
In the most of the previous studies, concatenated quantum error
correcting codes were considered instead of the surface codes
in the present study.

As discussed in detail in Ref.\cite{dennis,wang}, 
$p_c$ of the toric code can be obtained by studying the RBIM in
$2D$ and/or the $Z_2$ RPGM in $3D$. We note that $p_c$ depends on
the practical methods of error corrections which one employs. 
In the method of Ref.\cite{wang} it is estimated as $p_c = 0.029$.
Here we shall review the estimation of $p_c$ by using
the statistical-mechanical models in two steps.
In the first step, we assume that every measurement of check operators 
done in order to check the  quantum states is perfect, i.e., without 
any errors. Then the relevant model
to calculate $p_c$ is shown to be  the $2D$ RBIM. In the second step,
we take into account the possible errors in measuring check operators
themselves. In this case, the relevant model becomes  the $3D$ $Z_2$ RPGM. 
The latter model resembles
the $2D$ RPGM of the toric code discussed in the previous section, 
but the dimensions of two models are different. The third dimension
is introduced as the time direction along which we
measure check operators successively. 

Let us start the first step. There are two types of errors; 
(i) phase errors associated with fermion-pair states of (\ref{Fstate}) 
and (ii) spin-flip errors associated with vortex-pair states
of (\ref{Vstate}).  Since these two types of errors
can be discussed in the same manner,
we focus here on the phase errors. A typical examlpe
of phase-error syndrome is given in Fig.4.
The sites marked by filled circles denote error sites on which 
the check operator $A_x$ has the eigenvalue $(-1)$, while 
$A_x=1$ on the remaining sites.
This implies that a chain of the phase errors have occurred.
In Fig.5, we show one of the simplest case of two phase-error
sites, where the error chain containing all error links is denoted by $E$.
Therefore on the boundary sites of $E$, $A_x=-1$.
In order to correct these errors, we have to connect  
the error sites by a certain chain which we call $E'$, 
and apply a phase  operator
$(\prod Z_{xi})$ along $E'$ upon the quantum state under problem.
There is of course ambiguity in choosing a suitable $E'$.
Let us see this in detail. 
Because $E$ {\em plus} $E'$ form a closed
loop, i.e., a cycle, the corrected state and the original state 
differ by the factor $\prod_{E+E'}Z_{xi}$.
If  this closed loop belongs to the homologically trivial class 
of the set of  closed loops on the torus, we have
\begin{eqnarray}
\prod_{E+E'}Z_{xi} = \prod_{P\ {\rm with\ in}\ E+E'}B_P =1,
\label{correction}
\end{eqnarray}
beause of $Z_{xi}^2 =1$ and the last equation of Eq.(\ref{Grand}).  
Thus the correction is successful. 
On the other hand, if the closed loop belongs to the homologically
nontrivial class, 
the first equality of (\ref{correction}) does not hold, hence
the above correction procedure does not work.

Let us consider a toric code of an arbitrary large size and assume
that the probability of error per qubit is $p$.
An error chain $E$ is characterized by a function of each link $n_E(\ell)$ 
($\ell$ = link) like $n_E(\ell)=1$ for $\ell \in E$ and $n_E(\ell)=0$
for $\ell \notin E$.
As we explained above $E'=E+C$ where $C$ is a cycle (i.e., a chain with no
boundary) and then we define a function $n_C(\ell)$ for $C$ in the 
exactly same way with $n_E(\ell)$.
The probability that error chain $E$ occurs is given as
\begin{equation}
\mbox{prob}(E)=\prod_{\ell}(1-p)^{1-n_E(\ell)}p^{n_E(\ell)}.
\end{equation}
Let us consider the probability distribution for an
arbitrary chain $E'$ that has the same boundary with $E$, 
$\mbox{prob}(E'|E)$\footnote{The actual probability for the 
chain $E'$ is given by $\mbox{prob}(E)\cdot \mbox{prob}(E'|E)$.}.
For $\ell \in E'$ and $\ell \notin E$, the link functions are
$n_C(\ell)=1$ and $n_E(\ell)=0$ and its probability is given as
\begin{equation}
{p\over 1-p}.
\end{equation}
On the other hand for $\ell \notin E'$ and $\ell \in E$ with 
the link functions $n_C(\ell)=1$ and $n_E(\ell)=1$,
\begin{equation}
{1-p\over p}.
\end{equation}
For $n_C(\ell)=0$, the probability is unity.
Then the conditional probability is given by 
(up to an irrelevant constant factor)
\begin{equation}
\mbox{prob}(E)\cdot
\mbox{prob}(E'|E)= \prod_{\ell}\exp (J\eta_\ell u_\ell),
\label{pEE}
\end{equation}
where 
\begin{equation} 
u_\ell =1-2n_C(\ell)\in \{1,-1\},
\label{ul}
\end{equation}
\begin{equation}
\eta_\ell=
\left\{
\begin{array}{rl}
1, & \ell \notin E \\
-1, & \ell \in E,
\end{array}
\right.
\end{equation} 
and $e^{-2J}=p/(1-p)$.

From Eq.(\ref{ul}) and the fact that $C$ is a close chain,
\begin{equation}
\prod_{\ell\in x}u_\ell=1,
\label{ul2}
\end{equation} 
at each site $x$.
The above constraint (\ref{ul2}) can be easily solved by introducing classical
Ising spin variables $\sigma_{x^\ast}=\pm 1$ at sites of the dual lattice,
\begin{equation}
u_\ell =\sigma_{x^\ast}\sigma_{x^\ast+i},
\label{ul3}
\end{equation}
where the notation is obvious.
Then the generating function of the probability distribution of the error 
chains $E'$ is given by the following partition function $Z$ of the
RBIM,
\begin{equation}
Z[J,\eta]=\sum_{\{\sigma_{x^\ast}\}}\exp \Big(J\sum_\ell \eta_\ell 
\sigma_{x^\ast}\sigma_{x^\ast+i}\Big).
\label{Z}
\end{equation}
The relation $e^{-2J}=p/(1-p)$ between the temperature and the 
concentration of the ``wrong-sign" bonds $p$ 
defines the Nishimori line\cite{nishimori}. (This condition
of the Nishimori line will be expained in Sec.4.)
It is straightforward 
to express each contribution to the above partition function $Z$ 
pictorially in terms of the domain walls (see Fig.6).
From the above discussion, it is obvious that if the RBIM is in the
disordered phase, arbitrary large $E'$ appears and  successful
error corrections are impossible.
On the other hand, we can perform successful error corrections 
if the RBIM is in the ordered phase. 
The $2D$ RBIM has been studied well and its phase structure is known.
When errors and noises throughout the measurement
of the check operators are totally ignored,
the accuracy threshold of the toric quantum memory is given by the
intersection point of the phase boundary and the Nishimori line 
of the RBIM which is estimated as $p_c=0.109$\cite{RBIM}.

Now let us proceed to the second step by
taking into account the errors in measuring the check operators
themseves.
In this case, the error chains $E$ live in the two-dimensional space
and the one-dimensional time because we must repeat the check operations 
(syndrome measurements) many times.
Boundaries of the error chain $E$ reside on sites of the original $3D$
lattice.
In the $3D$ cubic lattice, a link of the original lattice corresponds to 
a plaquette of the dual lattice.
Then the constraint (\ref{ul2}) can be solved as 
\begin{equation}
u_\ell= \prod_{P^\ast}\sigma_{\ell^\ast},
\label{uell2}
\end{equation}
where the plaquette $P^\ast$ on the dual lattice corresponds to 
the link $\ell$ on the original lattice
and $\sigma_{\ell^\ast}$ is the Ising variable on the 
link $\ell^\ast$ of the dual lattice.
From Eqs.(\ref{pEE}) and (\ref{uell2}),
the conditional probability $\mbox{prob}(E'|E)$ can be expressed as 
\begin{equation}
\mbox{prob}(E)\cdot
\mbox{prob}(E'|E)= \prod_{P\ast}\exp(J\eta_{P^\ast}
\prod_{\ell^\ast\in P^\ast}\sigma_{\ell^\ast}),
\end{equation}
where $\eta_{P^\ast}=\eta_\ell=\pm 1$ and other notations are self-evident.
Then the statistical-mechanical model in the present case is 
the RPGM defined on the $3D$ lattice. Its third direction
represents the real time directions along which measurements are done,
while each plane lying in the $1-2$ directions is the dual lattice 
of the origina $2D$ lattice.

In the following section, we shall study the phase structure of the
RPGM by numerical methods.
Hereafter we assume that the error probability of syndrome measurement 
is also given by $p$ for simplicity.
We shall consider $3D$ lattice as large as possible.
As discussed in Refs.\cite{dennis,wang}, the phase structure of an arbitrary
large system gives the accuracy threshold of the toric code for the error 
correction.


\section{Phase structure of the RPGM}

In this section, we study the phase struture of the $3D$ $Z_2$
RPGM by MC simulations, and calculate the accuracy threshold for the
toric quantum memory. The phase boundary is determined by
calculating the specific heat and expectation values of the Wilson loop.
We note that our choice of the above model implies that 
we take into account possible errors which may
occur in the measurements of the check operators. 

The partition function of the $3D$ RPGM is given by
\begin{equation}
Z(\beta,\eta)=\sum_{\{\sigma_{xi}=\pm 1\}}e^{-\beta E}, \,\,\,
E=-\sum_{P}\eta_P  \prod_P \sigma,
\label{Z1}
\end{equation}
where $E$ is the energy and 
$\sigma_{xi} = \pm1$ is a $Z_2$ classical variable  
 sitting on the link $(x,i)$ of the $3D$ dual lattice 
$(i= 1,2,3)$.\footnote{As discussed in the previous sections,
the present RPGM is defined on the {\em dual} lattice.
In this section however, we shall use simple notations
like $P$ instead of $P^\ast$ for the dual plaquette, etc.}
$\prod_P \sigma$ denotes the product of four $\sigma$'s along 
the four links forming $P$.  
 $\eta_P=\pm 1$ is a {\em quenched} random variable on the plaquette $P$.
 The concentration
of ``wrong-sign" plaquettes with $\eta_P=-1$ is $p$ as in the previous 
sections.
Thus the weight function $P(K_P,\eta)$ for  a set of $\eta_P$'s is given by
\begin{eqnarray}
P(K_P,\eta)&=&\prod_P\left[(1-p)\ \delta_{\eta_P,1}+
p\ \delta_{\eta_P,-1}\right]
\nonumber\\
&=&(2\cosh K_P)^{-N_P} \cdot \exp 
\Big(K_P\sum_{P}\eta_P\Big),
\end{eqnarray}
where $N_P$ is the total number of plaquettes, and
the parameter $K_P$ is related with $p$ through the relations,
\begin{equation}
1-p \propto \exp(K_P),\ p \propto \exp(-K_P),\ {1-p \over p}=e^{2K_P}.
\label{KP}
\end{equation}
Then the  ensemble-averaged free energy $F(\beta, K_P)$ is defined as 
\begin{equation}
F(\beta,K_P)=-\frac{1}{\beta}\sum_{\{\eta_P=\pm 1\}}
P(K_P,\eta) \; \ln Z(\beta,\eta).
\label{JF}
\end{equation}

We presented in Fig.1 a schematic phase diagram of the RPGM
in the $p-T$ plane.
At $p=0$, i.e., the $Z_2$ gauge system in $3D$ with 
a constant gauge coupling, it is known that
there are two phases, one is the Higgs phase 
and the other is the confinement phase\cite{kogut}.
The critical coupling is estimated as $\beta_c=0.76 (T_c = 1.31)$.
As the concentration $p$ of ``wrong-sign" plaquettes increases,
the critical temperature $T_c(p)$ at $p$ decreases and 
at a certain value of $p=p_{0}$, the critical temperature 
vanishes, $T_c=0$. 
As discussed in the previous section, the system must be in the ordered 
Higgs phase  
for the toric code to work as a reliable quantum memory.
In Fig.1 the Nishimori line $T=T_{\rm N}(p)$ 
is plotted by the dashed line.
In the high-$T$ region {\em above} $T_{\rm N}(p)$, 
the critical line is rather horizontal,
whereas in the low-$T$ region {\em below} $T_{\rm N}(p)$,
it is almost vertical.

The Nishimori line is defined in the $p-T$ plane by setting
$K_P=\beta$, on which the quantum fluctuations of gauge field 
$\sigma_{xi}$ and fluctuations caused by the ``random impurity" 
$\eta_P$ valance in magnitude, i.e.,  giving the same effect
to disorder the system.
Let us explain this briefly.
For the case $p=0$, there is no randomness, and  
the quantum(thermal) fluctutions are contorolled by the
Botzmann factor $\exp(\beta \sum_P \prod \sigma)$.
On the other hand at $T = 0$ with finite $K_P$, 
the quantum fluctuations are suppressed;
$\sigma_{xi}$ are frozen and synchronized as $\prod_P \sigma = \eta_P$
due to the factor $\exp(\beta \eta_P \prod_P \sigma)$ in $Z$.
Thus the effects of randomness is controlled by the factor 
$P(K_P,\eta) \propto \exp(K_P  \sum_P \prod \eta)$.
These two factors become equal at $K_P = \beta$. 

Let us choose the statistical-mechanical models
as the method of error corrections in quantum computations. 
As discussed in Sec.3, we generate error-correcting chains $E'$
exactly in the same way with error chains $E$, and 
this procedure corresponds to the Nishimori line in the $p-T$ plane. 
The critical line and the Nishimori line intersect with each other
at $(p_c,T_{\rm N}(p_c))$. 
In order that the toric quantum codes to be fault-torelant,
the parameters of the toric codes, i.e., $(p,T)$, must be in the Higgs phase.
The intersection point $(p_c,T_{\rm N}(p_c))$ determines
the maximum possible value of $p$ for fault-tolerant codes, 
the accuracy threshold, as $p_c$.

Let us first focus on the high-$T$ region.
In Fig.7, we plot the specific heat per site, $c(\beta,p)$, as a function
of $T$ for various samples with $p=0.005, 0.01, 0.015, 0.02$ and $0.025$.
For $p=0.005$ the peak of the specific heat develops as we increases
the system size just as in the well studied case of $p=0$.
However, for $p=0.01$, the peak 
does not develop as shown in Fig.8.
For larger $p$'s, the peak of the specific heat disappears, 
although there seems to exist some
discontinuities in the derivative $dc/d\beta$. 
Below we shall examine these anomalous behaviors of $c(\beta,p)$
by calculating the expectation values of the Wilson loops, 
and conclude that there certainly exist
phase transitions at these values of $p$. 
The locations of these phase transitions are pointed
 with the arrows in Fig.7.  
The order of these transitions are weak second order
or  may be higher than 2.
In Fig.9 we plot the critical temperature $T_c(p)$
of these phase transitions
in the $p-T$ plane together with the solid Nishimori line.
$T_c(p)$ starts from $T_c(0)=1/0.76=1.31$ at $p=0$ 
and decreases as $p$ increases, as we expected.

To confirm the existence of the above phase transitions, 
we calculated the expectation values of the Wilson loop, 
\begin{eqnarray}
W(C)&=&\sum_{\{\eta_P\}}P(K_P,\eta)
\sum_{\{\sigma_{xi}\}}\Big(\prod_C \sigma_{xi}\Big)
\frac{e^S}{Z(\beta,\eta)} \nonumber \\
&=& [W_{\mbox{\footnotesize{sample}}}(C)]_{\mbox{\footnotesize{ens}}},
\label{wilson}  
\end{eqnarray}
\begin{equation}
W_{\mbox{\footnotesize{sample}}}(C)=
\sum_{\{\sigma_{xi}\}}\Big(\prod_C \sigma_{xi}\Big)
e^S/Z(\beta,\eta),
\label{sample}
\end{equation}
where $C$ denotes a closed loop in the $3D$ lattice
and $\prod_C \sigma_{xi}$ is the product of $\sigma_{xi}$'s
along $C$.  
$W_{\mbox{\footnotesize{sample}}}(C)$ is the expectation value of the Wilson
loop for {\em each sample} with fixed $\{\eta_P\}$ and 
$[{\cal O}]_{\mbox{\footnotesize{ens}}}$ denotes the ensemble average 
{\em over samples} of the quantity ${\cal O}$ calculated for
each sample.
$W(C)$ is introduced by Wilson to characterize the confinement
phase by the so called area law\cite{kogut}, that is, 
$W(C)$ behaves for large $C$'s as
\begin{eqnarray}
W(C) \propto \exp(-\alpha A(C)),
\label{area}
\end{eqnarray}
where $A(C)$ is the smallest area of all the branes 
whose boundaries are $C$, and $\alpha$ is a constant called the 
string tension. On the other hand, in the deconfinement phase 
like the Higgs phase, $W(C)$ exhibits the following perimeter
law for large $C$'s,
\begin{eqnarray}
W(C) \propto \exp(-\gamma P(C)),
\label{perimeter}
\end{eqnarray}
where $P(C)$ is the length of $C$.
In Fig.10 $W(C)$ are plotted for $p=0,0.01$ and $0.02$ from the above.
We chose a pair of $\beta$'s for each $p$, slightly above and 
below $T_c(p)$ that are determined by the calculation of 
the specific heat (see Fig.7).
In the left column of Fig.10, $W(C)$ are plotted as  
$-\ln(W(C))/A(C)$ vs $A(C)$, so the curves become constant
$\alpha$  when the area  law (\ref{area}) holds.
In the right column of Fig.10, $W(C)$ are plotted as  
$-\ln(W(C))/P(C)$ vs $P(C)$, so the curves become constant
$\gamma$  when the perimeter  law (\ref{perimeter}) holds.
At all the pairs of $\beta$'s presented in the figures,
we observe the changes in the behavior of $W(C)$ from 
the perimeter law at $T<T_c(p)$ to the area law at $T_c(p)<T$.
These changes take place at temperatures very close to $T_c$, 
which confirm the existence of the phase transitions.
The obtained value of $T_c(p)$ by the Wilson loop is {\em insensitive} to the 
distribution of the wrong-sign plaquettes, i.e., $\{\eta_P\}$
for fixed $p$.

Next we study the low-$T$ region.
We calculated the specific heat per site $c(\beta,p)$ 
along a line of fixed $T$ by changing $p$, i.e., concentration
of the wrong-sign plaquettes.
For each $p$, we prepared 1000 samples, each of which has
a definite configuration of $\{\eta_P\}$, and calculated the specific
heat for each sample and averaged over the results.
To obtain the suitable initial configuration of $\{\sigma_{xi}\}$
for each sample at low $T$,
we employed the quenched-annealing method.
In Fig.11, the averaged values of the specific heat for $\beta=2.5$  
are plotted from $p=0.005$ to $p=0.06$.
We superimpose the fluctuations of each sample around the average.
There exists a samll discontinuity in $c(\beta=2.5,p)$ at
$p=0.027 \sim 0.028$.
As $p$ increases, the fuctuations also reduce suddenly at
$p=0.027 \sim 0.028$ and they become almost constant for
$p>0.032\sim 0.033$, so some kind of transition seems to occur there.
As we show shortly, the same behavior of the specific heat 
is observed at $\beta=2.0$.
In order to verify the above expectation, we calculated the 
expectation values of the Wilson loop, $W(C)$ of (\ref{wilson}).
More precisely, we first calculated expectation value of the Wilson 
loop $W_{\mbox{\footnotesize{sample}}}(C)$
for each sample with {\em fixed} $\{\eta_P\}$.
Typical results are shown in Fig.12 for $p=0.013,0.024,0.025$ and $0.030$.
It is obvious that a change of behavior of 
$W_{\mbox{\footnotesize{sample}}}(C)$ takes place
from the perimeter law (\ref{perimeter}) to the area law (\ref{area}) 
as $p$ increases as we expected.
In the region close to $p=0.026$, some samples show the area law,
whereas others exhibit the perimeter law.
This result is in contrast to that of the high-$T$ region.

In order to understand the implication of the above result, 
let us see the expectation value of the Wilson loop more precisely.
From Eq.(\ref{wilson}), it is obvious that 
the samples with the perimeter law dominate over those with
the area low as $e^{-P(C)}\gg e^{-A(C)}$ for large $C$
and $W(C)$ exhibits the {\em perimeter law} if 
samples with the perimeter and area laws {\em coexist}.
We then examined $W_{\mbox{\footnotesize{sample}}}(C)$
for various but a finite number of samples by varying $p$ and 
found that certain samples at $p=0.025$ and $0.026$ exhibit the perimeter law
and others show the area law, i.e., the coexistence
(we show the result of a sample with 
the area law for $p=0.025$ in Fig.12). 
For larger values of $p$, we found only the area-law decaying 
$W_{\mbox{\footnotesize{sample}}}(C)$.
(This does not mean that all samples exhibit the area law because
we only examined a finite number of samples.)
In this way we estimate a lower bound of the critical value of $p$ as 
$p\sim 0.026$, which is consistent with the calculations of
the specific heat given in Fig.11.

We also studied the case of $\beta=2.0$ which is close to
the Nishimori line.
In Fig.13, the specific heat and the fluctuations over samples
are shown.  
On the Nishimori line, it is proved\cite{nishimori} that 
the specific heat exhibits {\em no} singular behaviors
(no divergences).
In fact, Fig.13
shows rather smooth
behavior in $c(\beta,p)$.
Let us  examine the possibility of  a phase transition by studying $W(C)$. 
In Fig.14 we present $W(C)$ at $\beta=2.0$ for 
$p=0.020,0.030,0.031$ and $0.040$. 
As $p$ increases, we observe that the behavior of $W(C)$ changes 
from the perimeter to area law.
As far as we observed,
the maximum value of $p$ at which $W(C)$ exhibits the perimeter law 
is $0.030$, and therefore we conclude that
the phase transition from the Higgs phase to the confinement phase 
occurs near this value.

From the calculations of the specific heat and its fluctuations
shown in Figs.11 and 13, we estimate the critical values of $p$ 
as $p=0.032 \sim 0.033$ at both $\beta=2.5$ and $\beta=2.0$.
In both cases, the fluctuations of the specific heat become constant
for $p>0.032 \sim 0.033$.
For the case of $\beta=2.5$, if we regard the samll discontinuity
in the specific heat as the phase transition point, the critical
value of $p$ is estimated as $\sim 0.027$.
In order to obtain a definitive value of $p_c$, a thorough investigation
on the Wilson loop is required.

\section{Conclusion}

In this paper, we studied the RPGM in $3D$ numerically and 
obtained the phase transition line in the $p-T$ plane which
separates the Higgs and confinement phases.
The critical concentration of the ``wrong-sign" plaquettes
is estimated as $p=0.032 \sim 0.033$ at low $T$ and also
close to the Nishimori line.
This result determines the accuracy threshold for the quantum toric
code as $p_c \simeq 0.033$. 
In Ref.\cite{wang}, the critical value of $p$ along the $T=0$ axis
was estimated as 0.029, which corresponds $p_{0}$ in Fig.1. 
This value gives a lower bound of $p_c$, and 
Fig.9 shows that our MC simulations 
are consistent with their calculations. 

In Ref.\cite{AI} a qualitative phase diagram of the $3D$ $Z_2$ RPGM was
obtained by using the replica methods, and 
the existence of the ``gauge-glass phase"
is predicted at relatively low $T$ and moderate value of $p$.
In the present calculations, however, there appears no sign of
that phase.
In the higher-dimensional gauge models, the gauge-glass phase
may exist.

It was recently suggested that anyonic excitations in Kitaev's
model with discrete non-Abelian gauge groups may play an important role
in constructing quantum gates of QC's\cite{mochon,benoit}.
Then generalization of the present study of the $Z_2$ model
to a non-Abelian gauge model is interesting.


\begin{figure}[htbp]
    \psfrag{T}[t]{$T$}
    \psfrag{P}[t]{$p$}
    \psfrag{C}[t]{$p_{0}$}
    \psfrag{I}[t]{$(p_c,T _{\rm N}(p_c))$}
    \psfrag{O}[t]{$0$}
    \psfrag{N}[t]{Nishimori}
    \psfrag{S}[t]{Line}
    \psfrag{H}[t]{high-$T$ region}
    \psfrag{L}[t]{low-$T$ region}
    \begin{center}
        \includegraphics{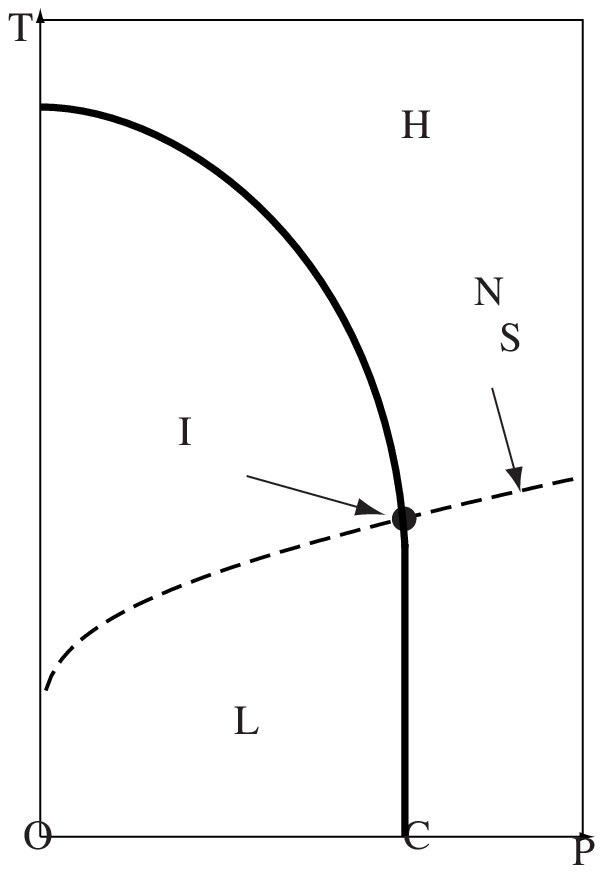}
    \end{center}
  \caption{A schematic phase diagram of the $3D$ $Z_2$ RPGM 
  in the $p-T$ plane,
  where $p$ is the concentration of wrong-sign plaquettes 
  and $T=\beta^{-1}$ is the ``temperature"
  (the gauge coupling constant). The thick curve is the phase-transition
  line $T_{c}(p)$ which separates the ordered Higgs phase and the 
  disordered confinement phase. The dashed line $T_{\rm N}(p)$
  is the Nishimori line which corresponds to the present methods of the error
  corrections in quantum computations. The value $p_c$ at the 
  intersection point $(p_c,T_{\rm N}(p_c))$ determines the 
  accuracy threshold for the quantum toric code.}
  \label{fig:01expected}
\end{figure}

\begin{figure}[htbp]
\begin{center}
\begin{tabular}{cc}
\subfigure[]{
\begin{minipage}{6.5cm}
\begin{center}
    \psfrag{A}[t]{$C^2_Z$}
    \psfrag{B}[t]{$C^1_Z$}
        \includegraphics{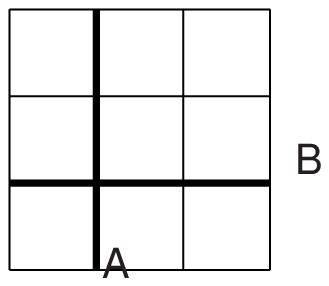}
\end{center}
\end{minipage}}
&
\subfigure[]{
\begin{minipage}{6.5cm}
\begin{center}
    \psfrag{A}[t]{$\tilde{C}^1_X$}
    \psfrag{B}[t]{$\tilde{C}^2_X$}
    \includegraphics{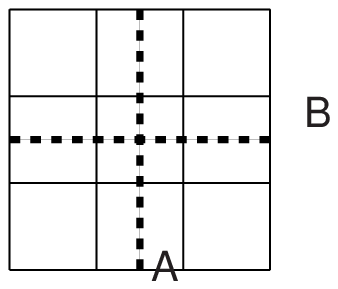}
\end{center}
\end{minipage}}
\end{tabular}
\end{center}
\caption{The original lattice (a) with periodic boundary
condition and its dual lattice (b).
$C^1_Z, C^2_Z, $ are two noncontractible closed loops on (a)
in the horizontal and vertical directions,
while $\tilde{C}^1_X,\tilde{C}^2_X $ are corresponding loops
on the dual site.
}
\label{fig:lattice}
\end{figure}

\begin{figure}[htbp]
  \begin{center}
    \psfrag{x}[t]{$x^*$}
    \psfrag{y}[t]{$y^*$}
    \includegraphics{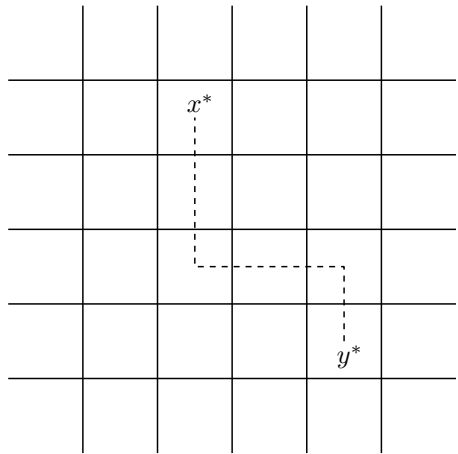}
  \end{center}
  \caption{Path $\tilde{C}_{x^* y^*}$ connecting dual sites $x^*$ and $y^*$.}
  \label{fig:03fig2}
\end{figure}

\begin{figure}[htbp]
  \begin{center}
    \psfrag{Z}[t]{$Z$}
    \includegraphics{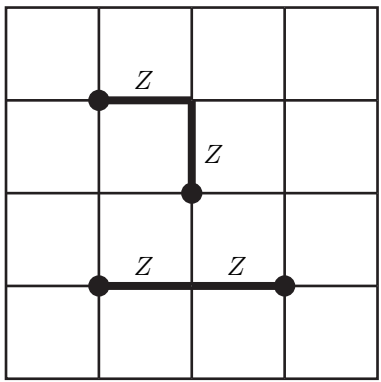}
  \end{center}
  \caption{Typical example of error syndrome produced by noises. 
  The error sites are
  marked by filled circles, where the check operators are $A_x=-1$.
  They are connected by the  error links marked 
  by thick links, on which extra $Z_{xi}$'s
     are applied. }
  \label{fig:04error_syndrome}
\end{figure}

\begin{figure}[htbp]
  \begin{center}
    \psfrag{E}[t]{$E$}
    \psfrag{F}[t]{$E'$}
        \includegraphics{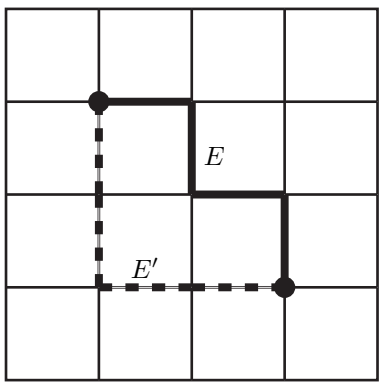}
  \end{center}
  \caption{An error chain $E$ produced by noises and an  applied chain $E'$ 
 produced in the process of the error correction. 
 Both $E$ and $E'$ connect the pair of
 error sites and $E+E'$  form a closed loop, i.e., a cycle.
 }
  \label{fig:cycle}
\end{figure}

\begin{figure}[htbp]
    \psfrag{E}[t]{$E$}
    \psfrag{F}[t]{$E'$}
    \begin{center}
        \includegraphics{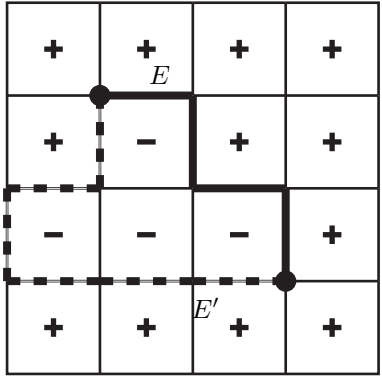}
    \end{center}
  \caption{Domain wall is produced along the cycle 
  of the error  chain $E$ and the error-correction chain $E'$.
  Corresponding configuration in the partition
  function (\ref{Z}) is achieved by the configuration
  of the dual variables,  $\sigma_{x^*} = -1$ 
  within the wall and $\sigma_{x^*} = +1$ 
  outside the wall.
   }
  \label{fig:06domain_walls}
\end{figure}

\begin{figure}[htbp]
    \begin{center}
        \psfrag{A}[t]{$\beta=0.788 (T=1.269)$}
        \psfrag{B}[t]{$\beta=0.822 (T=1.217)$}
        \psfrag{C}[t]{$\beta=0.866 (T=1.155)$}
        \psfrag{D}[t]{$\beta=0.900 (T=1.111)$}
        \psfrag{E}[t]{$\beta=0.990 (T=1.010)$}
        \psfrag{F}[t]{$\beta (=1/T)$}
        \includegraphics[scale=1.3]{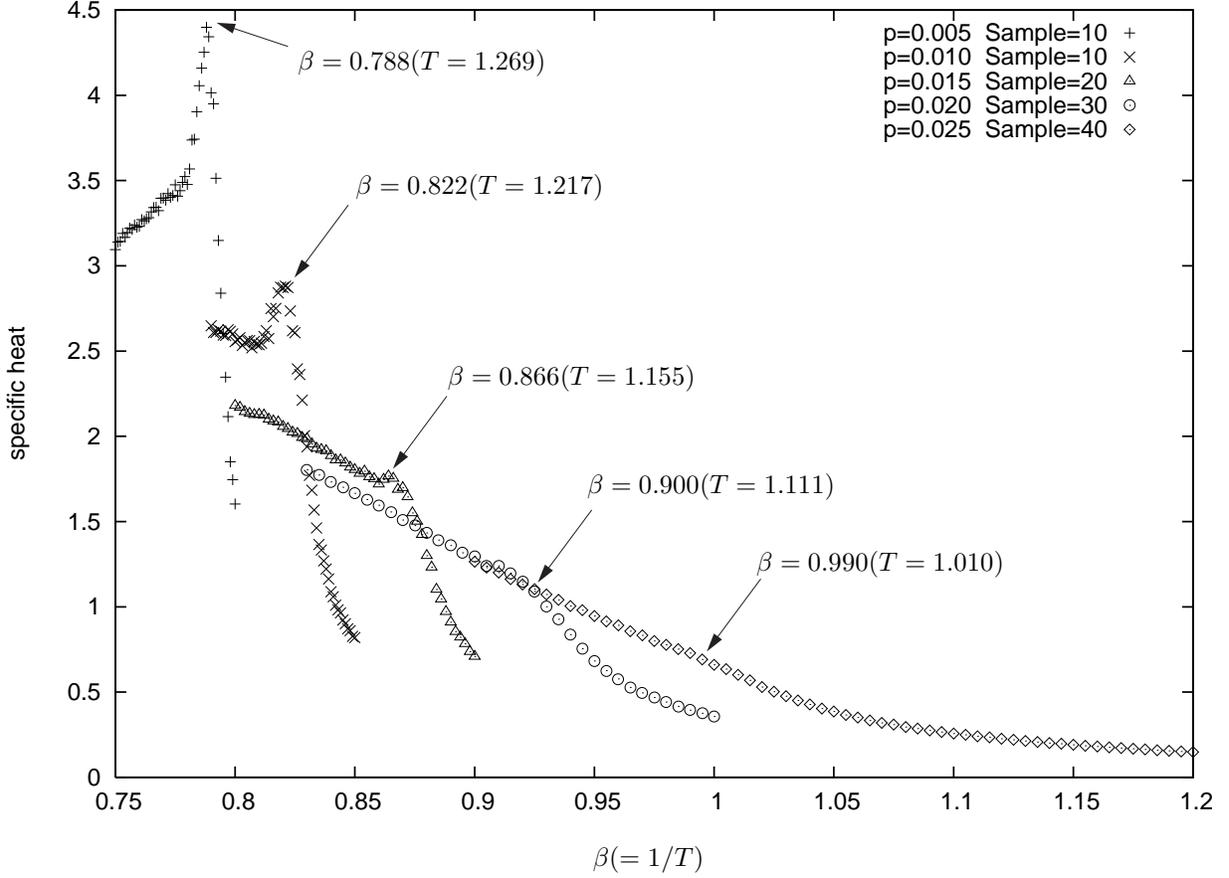}
    \end{center}
  \caption{Specific heat per site in the  high-$T$ region 
  vs $\beta$. In the MC simulation, the size of the $3D$ lattice 
  is $N_v=24^3$ sites, and the numbers of sweeps are 
  20000 for thermalization and 100000 for measurement for $p$=0.005, 
  0.010 and 0.015, and 30000+100000 for $p$=0.020 and 0.025.
  We averaged over $10\sim40$ samples with different configurations
  of $\eta_P$. 
  The positions marked by arrows denote the locations of phase 
  transitions confirmed by measuring the Wison loops.
 }
  \label{fig:07highT}
\end{figure}

\begin{figure}[htbp]
  \begin{center}
    \psfrag{C}[t]{{\scriptsize $N_v=20^3$}}
    \psfrag{D}[t]{{\scriptsize $N_v=24^3$}}
    \psfrag{B}[t]{$\beta(=1/T)$}
    \includegraphics{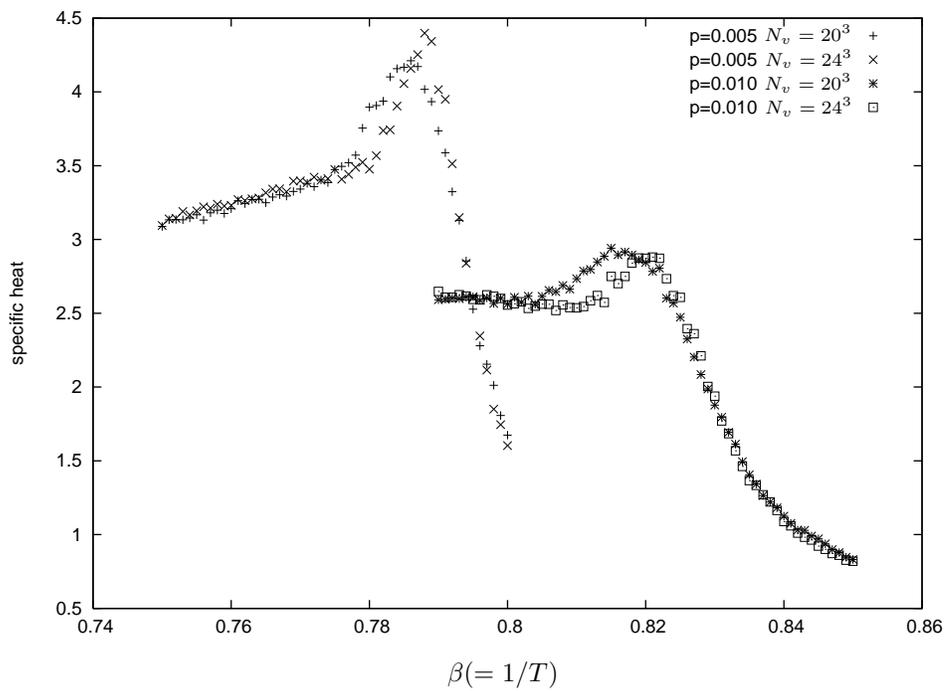}
  \end{center}
  \caption{Specific heat vs $\beta$ for $p=0.005, 0.01$.
  For $p=0.005$, the peak develops as the system size increases, 
  exhibitting a second-order
  transition. For $p=0.01$, the size dependence is very weak and the order of
  the transition seems to be very weak second order or
  higher than second-order.
  }
  \label{fig:08higher}
\end{figure}

\begin{figure}
    \begin{center}
        \psfrag{P}[t]{$p$}
        \psfrag{T}[t]{$T$}
        \includegraphics{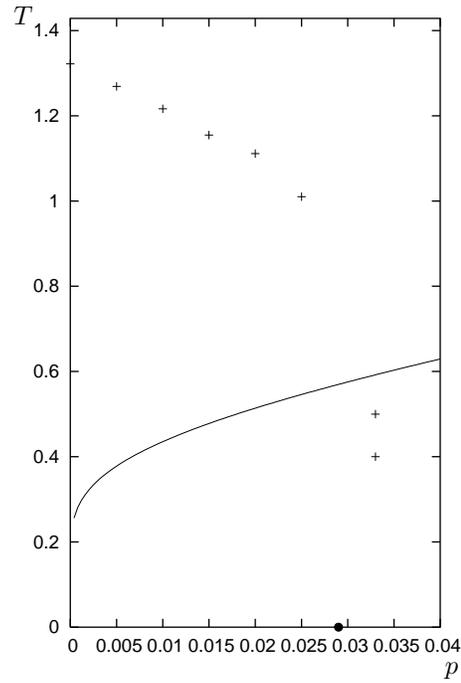}
    \end{center}
  \caption{Phase diagram in the $p-T$ plane. The crosses
  denote the phase transition points $T_c(p)$ determined by
  the specfic heat and the Wilson loops. The solid line
  is the Nishimori line $T_{\rm N}(p)$.
  As $T$ is lowered, the critical value of $p$
  starts to increase, and then decreases. This bending 
  behavior of $T_c(p)$ makes the problem of finding
  the best method of error corrections nontrivial.
  The statistical-mechanical methods of the error
correction discussed in Sec.3
  correspond to the Nishimori line, which gives a higher accuracy
  threshold $p_c \simeq 0.033$ than $p_0 =0.029$ 
  determined by $T_{c}(p_0)=0$\cite{wang}. 
  }
  \label{fig:09transition_line}
\end{figure}

\begin{figure}[htbp]
\begin{tabular}{ccc}
$p=0$
&
\subfigure[$p=0$, Area Law]{
\begin{minipage}{6.5cm}
    \psfrag{A}[t]{{\tiny $A$:Area}}
    \psfrag{Y}[t]{{\tiny $-\ln (W.L.)/A$}}
    \includegraphics[scale=0.5]{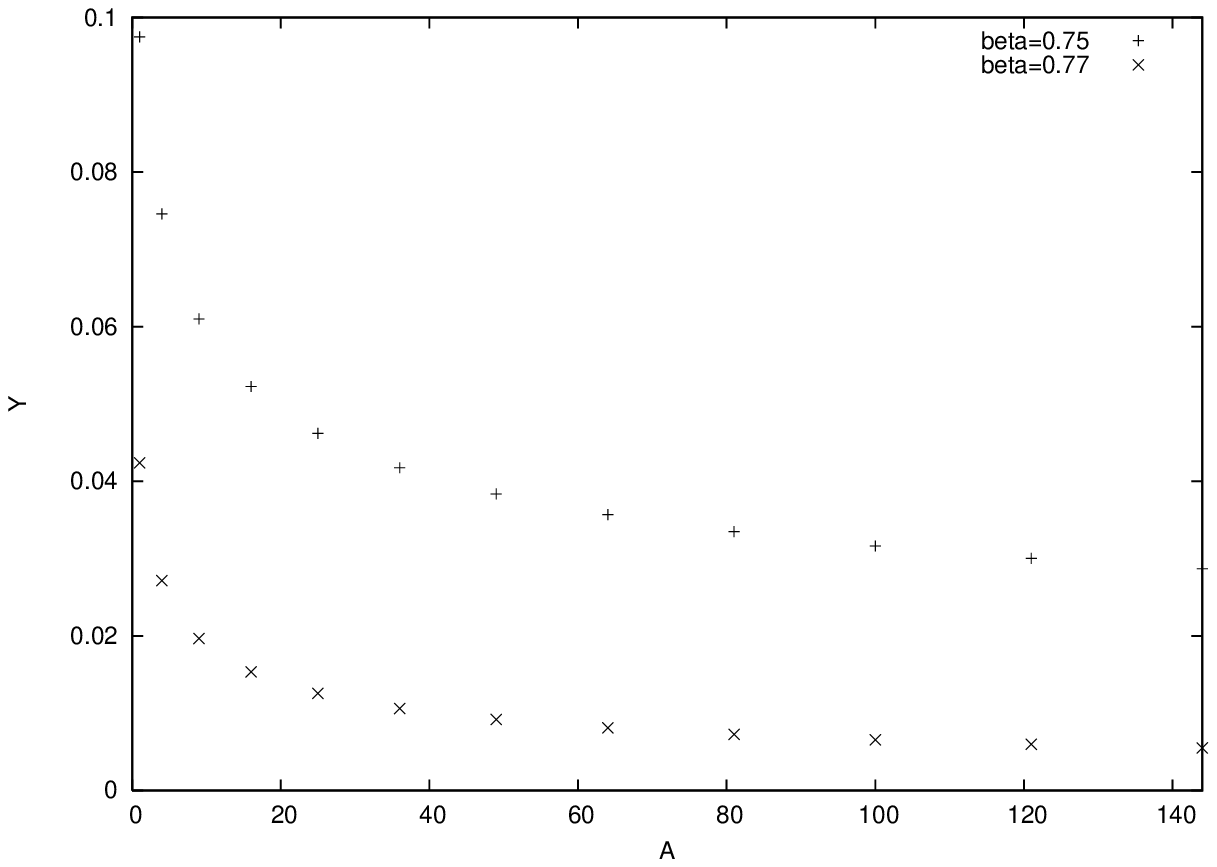}
\end{minipage}}
&
\subfigure[$p=0$, Perimeter Law]{
\begin{minipage}{6.5cm}
    \psfrag{P}[t]{{\tiny $P$:Perimeter}}
    \psfrag{Y}[t]{{\tiny $-\ln (W.L.)/P$}}
    \includegraphics[scale=0.5]{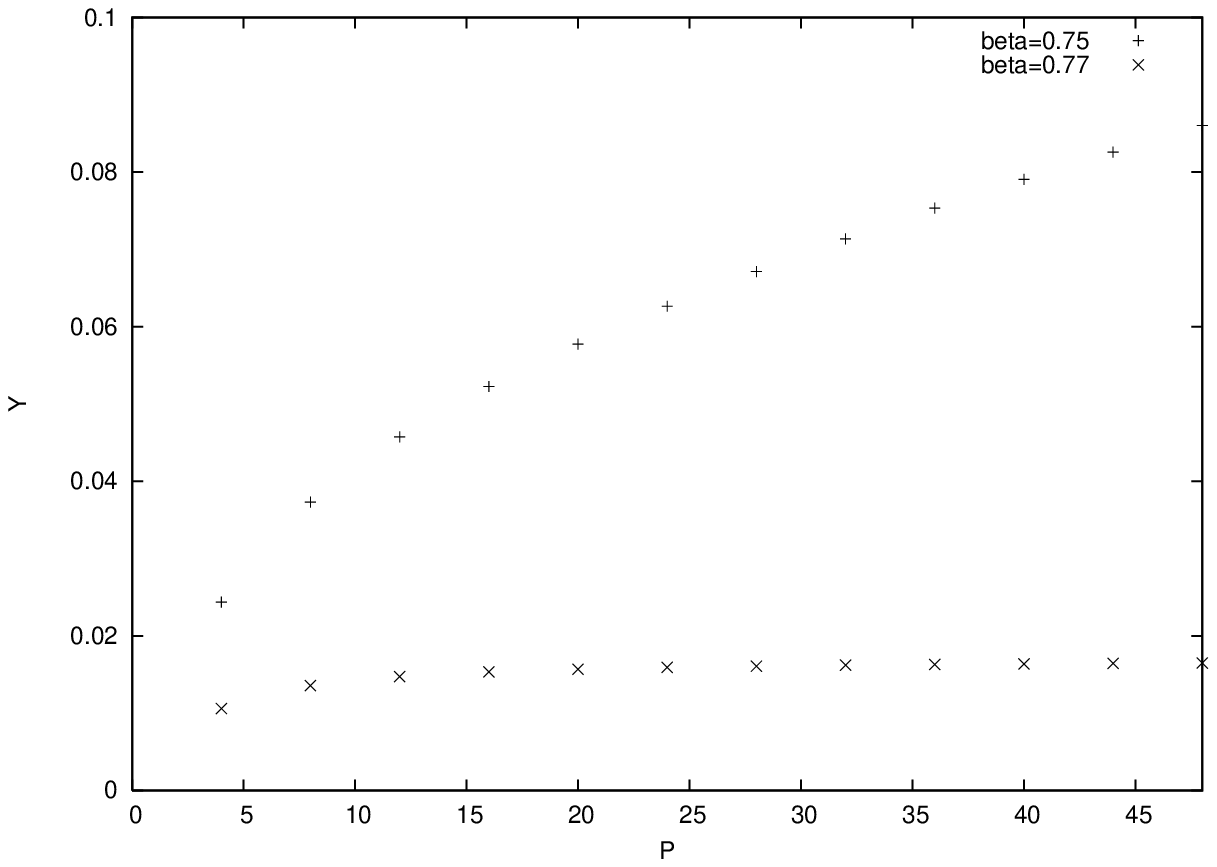}
\end{minipage}\vspace{5mm}}
\\
$p=0.010$
&
\subfigure[$p=0.010$, Area Law]{
\begin{minipage}{6.5cm}
    \psfrag{A}[t]{{\tiny $A$:Area}}
    \psfrag{Y}[t]{{\tiny $-\ln (W.L.)/A$}}
    \includegraphics[scale=0.5]{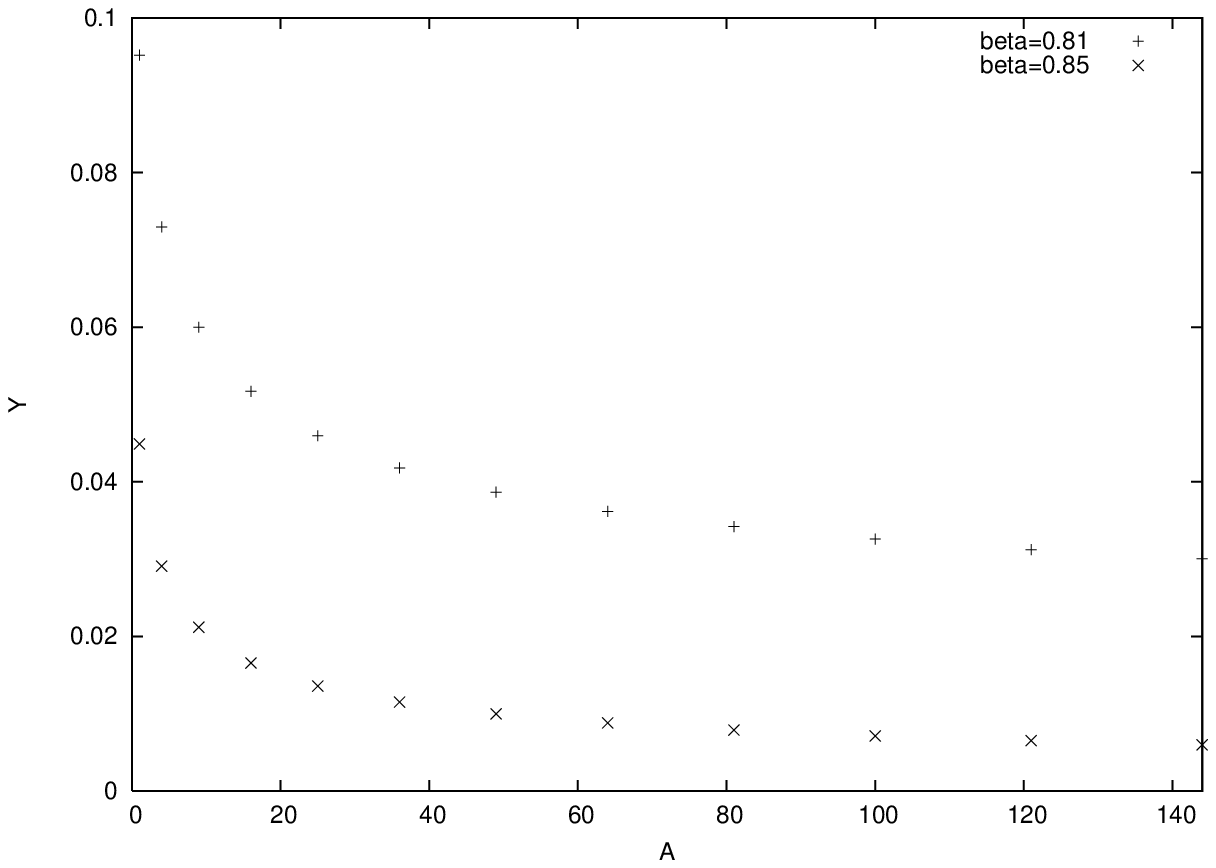}
\end{minipage}}
&
\subfigure[$p=0.010$, Perimeter Law]{
\begin{minipage}{6.5cm}
    \psfrag{P}[t]{{\tiny $P$:Perimeter}}
    \psfrag{Y}[t]{{\tiny $-\ln (W.L.)/P$}}
    \includegraphics[scale=0.5]{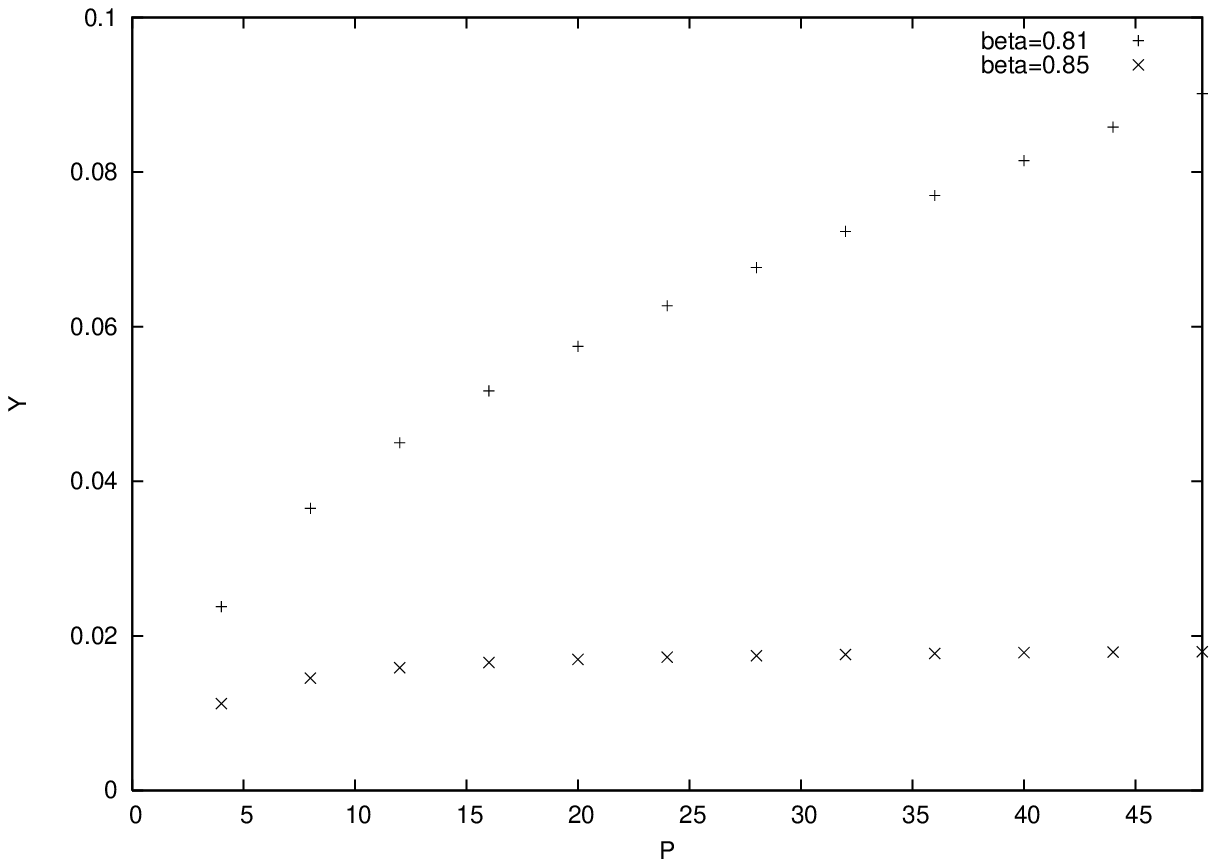}
\end{minipage}\vspace{5mm}}
\\
$p=0.020$
&
\subfigure[$p=0.020$, Area Law]{
\begin{minipage}{6.5cm}
    \psfrag{A}[t]{{\tiny $A$:Area}}
    \psfrag{Y}[t]{{\tiny $-\ln (W.L.)/A$}}
    \includegraphics[scale=0.5]{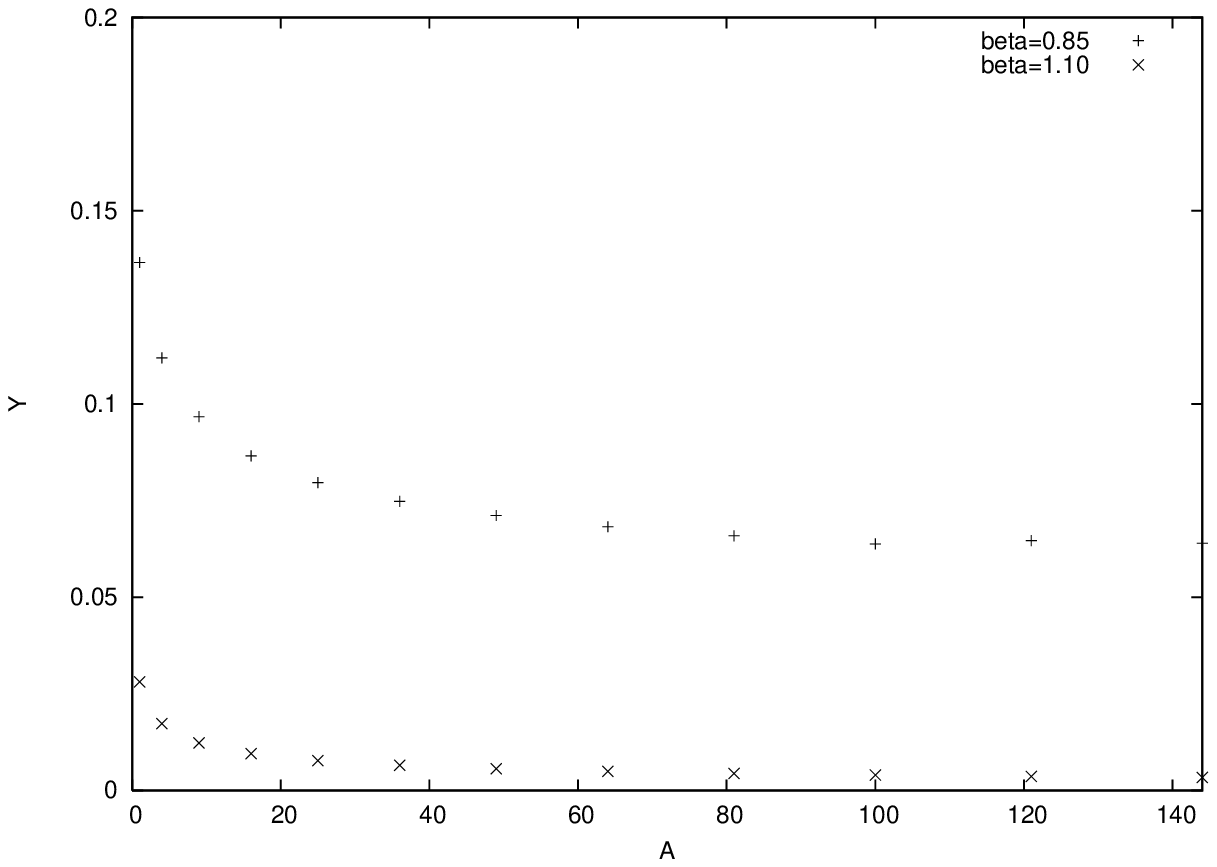}
\end{minipage}}
&
\subfigure[$p=0.020$, Perimeter Law]{
\begin{minipage}{6.5cm}
    \psfrag{P}[t]{{\tiny $P$:Perimeter}}
    \psfrag{Y}[t]{{\tiny $-\ln (W.L.)/P$}}
    \includegraphics[scale=0.5]{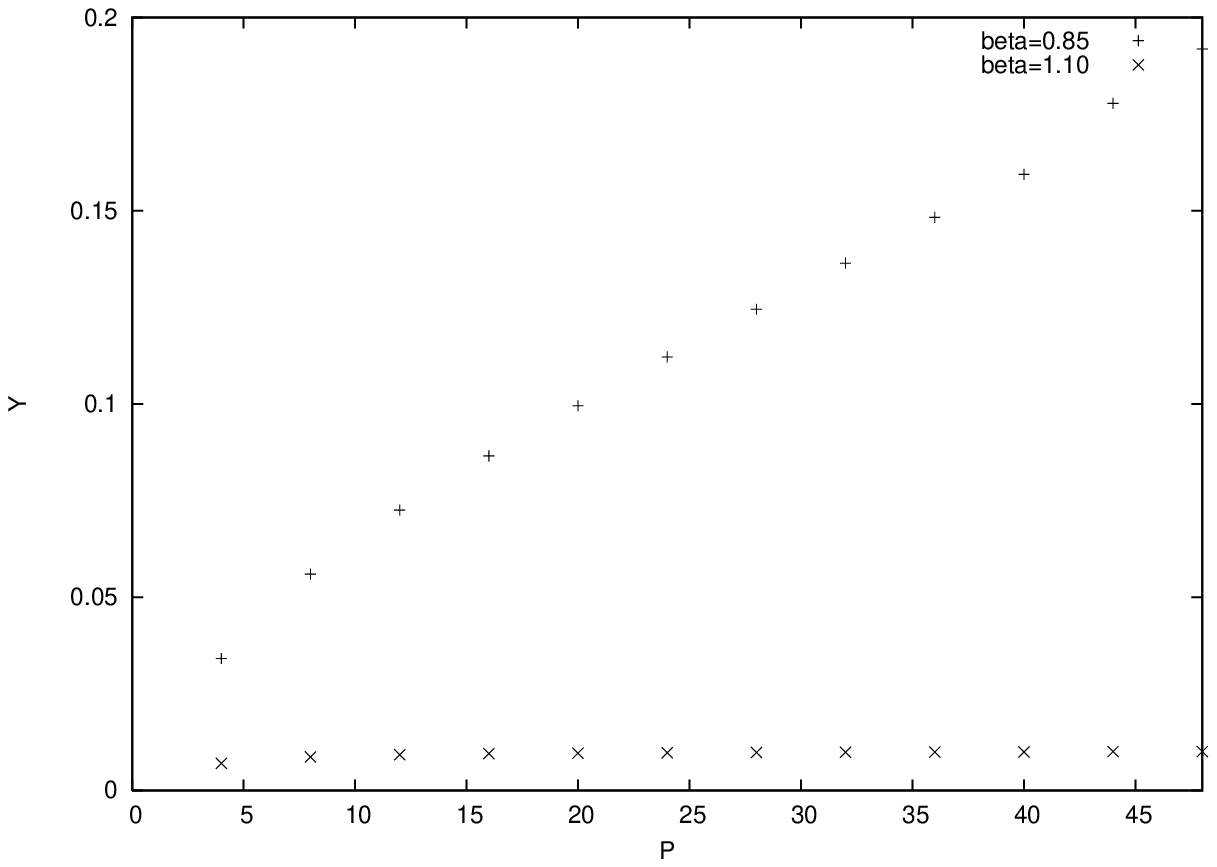}
\end{minipage}}
\end{tabular}
\caption{Wilson loops $W(C)$ in high-$T$ region.
The size of the lattice is $N_v=24^3$.
The results are plotted in the pair of axes,
$A(C)$(the smallest number of the plaquettes in all 
the branes bounded
by the loop $C$) and $-\ln(W(C))/A(C)$
in the left column,  and 
 $P(C)$(the number of links contained in $C$)
 and $-\ln(W(C))/P(C)$ in the right column.  
If the area(perimeter) law (\ref{area})((\ref{perimeter})) 
holds, the curves in the left(right) 
column become constants $\alpha(\gamma)$.
 }
 \label{fig:WL_in_highT_region}
\end{figure}

\begin{figure}[htbp]
\begin{center}
\psfrag{A}[t]{{\scriptsize $p=0.032$}}
\psfrag{B}[t]{{\scriptsize $p=0.033$}}
\psfrag{P}[t]{{\scriptsize $p$}}
    \begin{center}
        \includegraphics{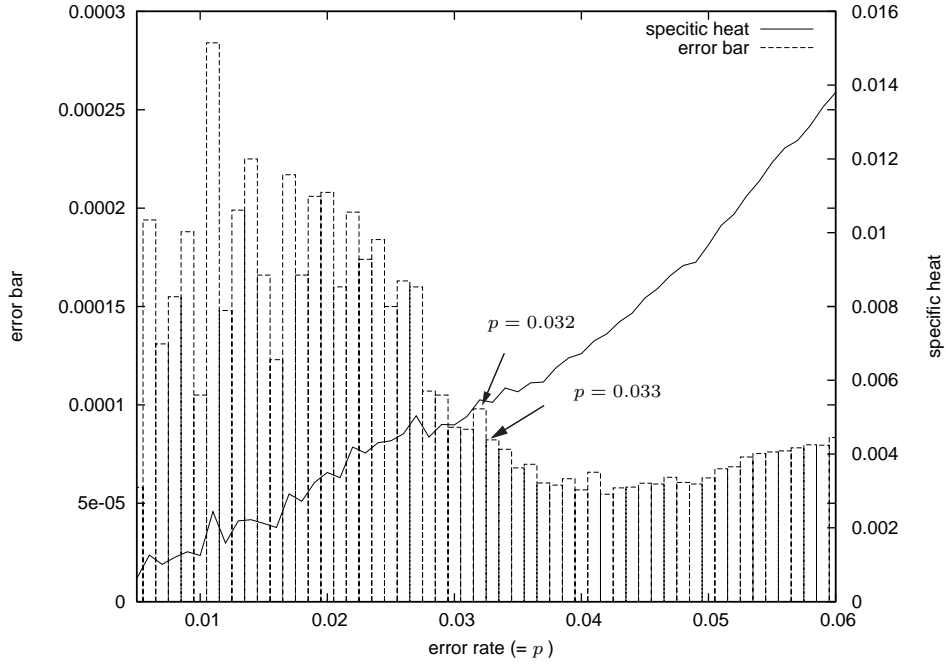}
    \end{center}
  \label{fig:11lowTbeta=2_5}
\end{center}
\caption{Specific heat in the low-$T$ region at $\beta=2.5$ vs $p$.
The histogram denotes the fluctuations at each $p$ over 1000 samples around
the averge. As $p$ increases, the fluctuation decreases and suddenly
reduces to an almost constant value at $p \simeq 0.033$ marked by arrows. 
This value
almost coincides with the critical value determined by $W(C)$
in Fig.12.
  }
\label{fig:lowTbeta=2_5}
\end{figure}

\begin{figure}[htbp]
\begin{tabular}{ccc}
$\beta=2.5$
&
\subfigure[$\beta=2.5$, Area Law]{
\begin{minipage}{6.5cm}
    \psfrag{A}[t]{{\tiny $A$:Area}}
    \psfrag{Y}[t]{{\tiny $-\ln (W.L.)/A$}}
    \includegraphics[scale=0.5]{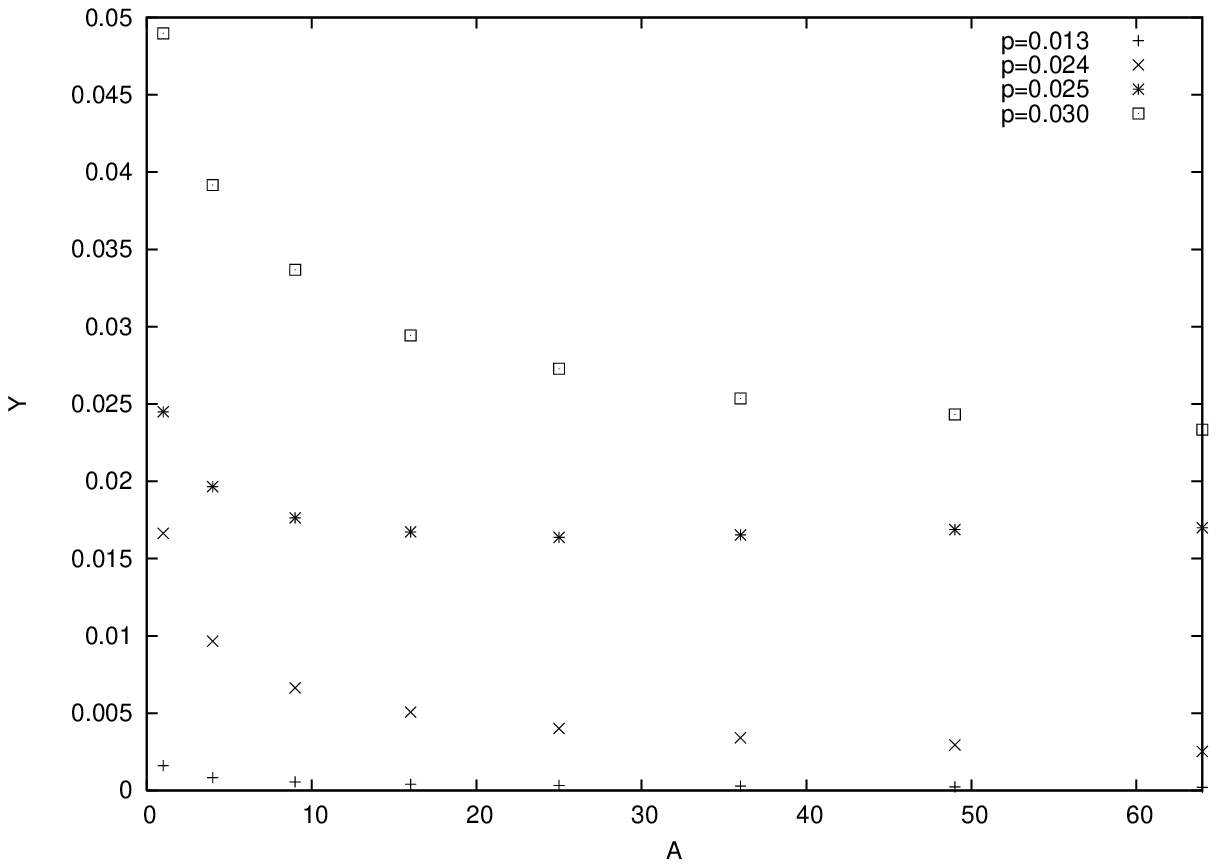}
\end{minipage}}
&
\subfigure[$\beta=2.5$, Perimeter Law]{
\begin{minipage}{6.5cm}
    \psfrag{P}[t]{{\tiny $P$:Perimeter}}
    \psfrag{Y}[t]{{\tiny $-\ln (W.L.)/P$}}
    \includegraphics[scale=0.5]{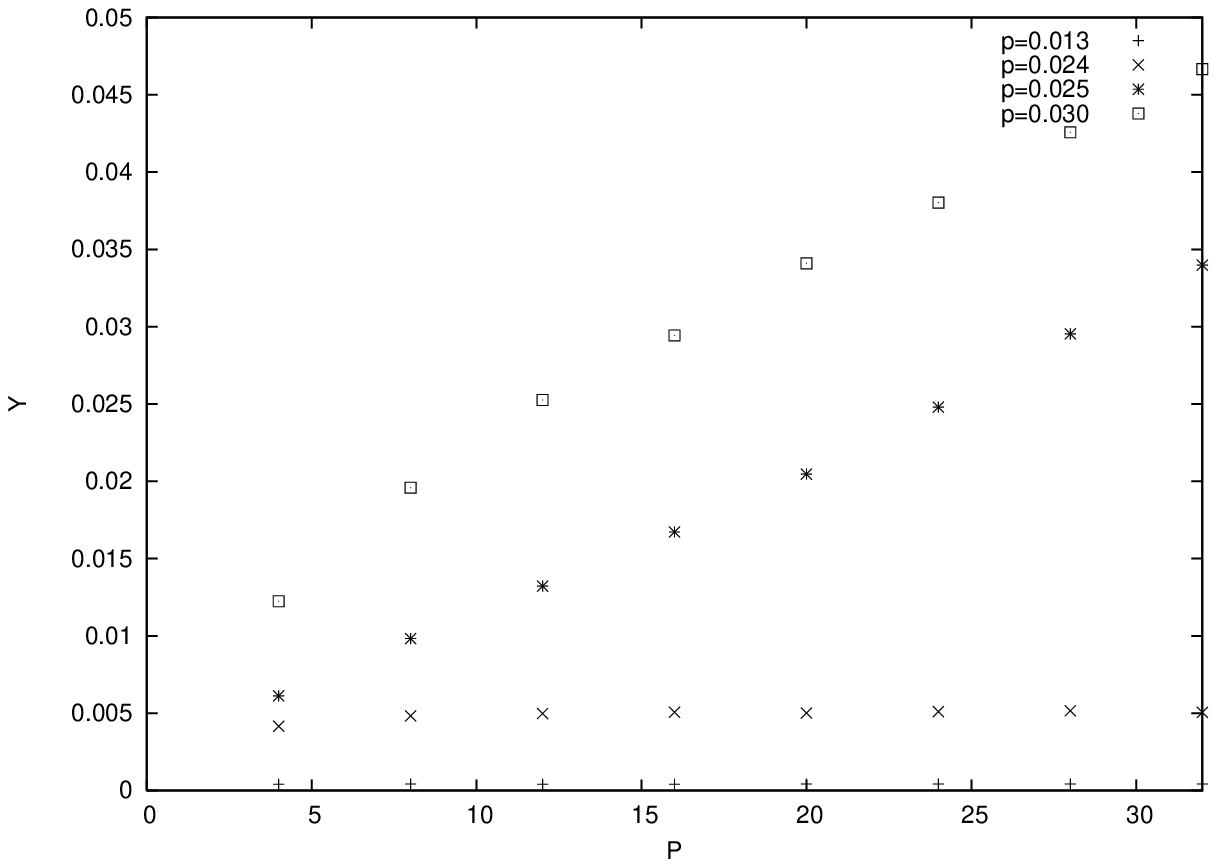}
\end{minipage}}
\end{tabular}
\caption{Wilson loop $W_{\mbox{\footnotesize{sample}}}(C)$
for  certain samples in the low-$T$ region  
at $\beta=2.5$.
The size of the lattice is $N_v=16^3$.
At $p= 0.025$ it exhibits the area law, while
at $p= 0.024$, the perimeter law.
This result gives a lower-bound estimate of the value of $p$ on criticality,
which is consistent with the specific-heat calculations in Fig.11. 
(See text.)
}
\label{fig:WL_in_lowT_region_beta=2.5}
\end{figure}

\begin{figure}[htbp]
\psfrag{A}[t]{$p=0.032$}
\psfrag{B}[t]{$p=0.033$}
\psfrag{P}[t]{$p$}
    \begin{center}
        \includegraphics[scale=1]{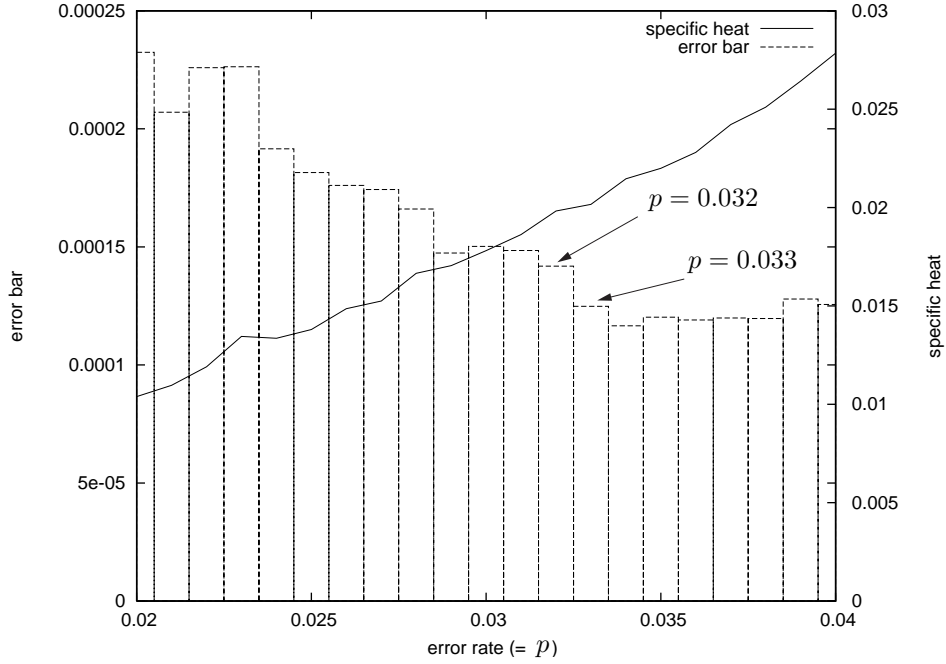}
    \end{center}
  \caption{Specific heat and fluctuations at $\beta=2.0$.
  Similar plot as in Fig.11. As $p$ increases, the 
  fuctuation reduces gradually, and reaches at 
  a constant value at $p\simeq 0.033$ as marked with arrows. 
  This value almost coincides with the critical value
  studied by the Wilson loops in Fig.14. 
  }
  \label{fig:13lowTbeta=2_0}
\end{figure}

\begin{figure}[htbp]
\begin{tabular}{ccc}
$\beta=2.0$
&
\subfigure[$\beta=2.0$, Area Law]{
\begin{minipage}{6.5cm}
    \psfrag{A}[t]{{\tiny $A$:Area}}
    \psfrag{Y}[t]{{\tiny $-\ln (W.L.)/A$}}
    \includegraphics[scale=0.5]{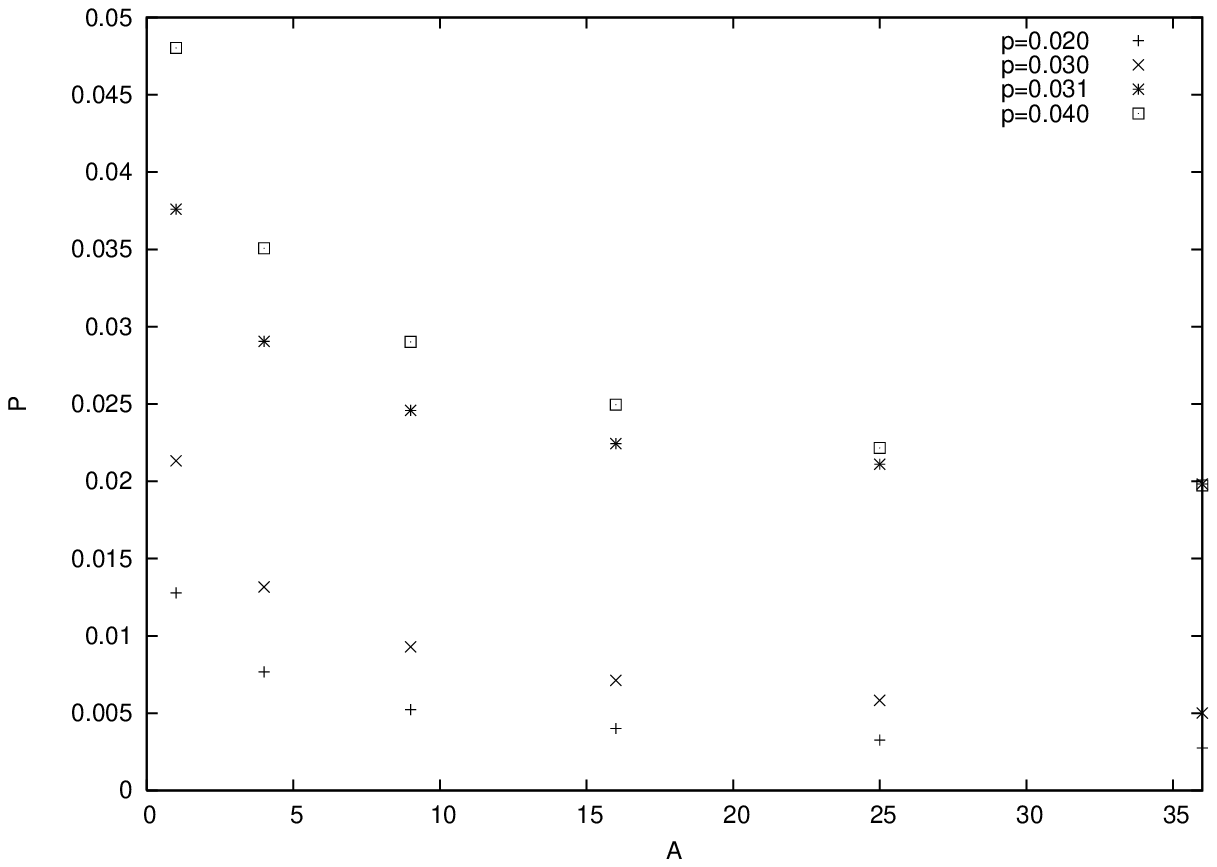}
  \label{fig:14_1lowTbeta=2_0A}
\end{minipage}}
&
\subfigure[$\beta=2.0$, Perimeter Law]{
\begin{minipage}{6.5cm}
    \psfrag{P}[t]{{\tiny $P$:Perimeter}}
    \psfrag{Y}[t]{{\tiny $-\ln (W.L.)/P$}}
    \includegraphics[scale=0.5]{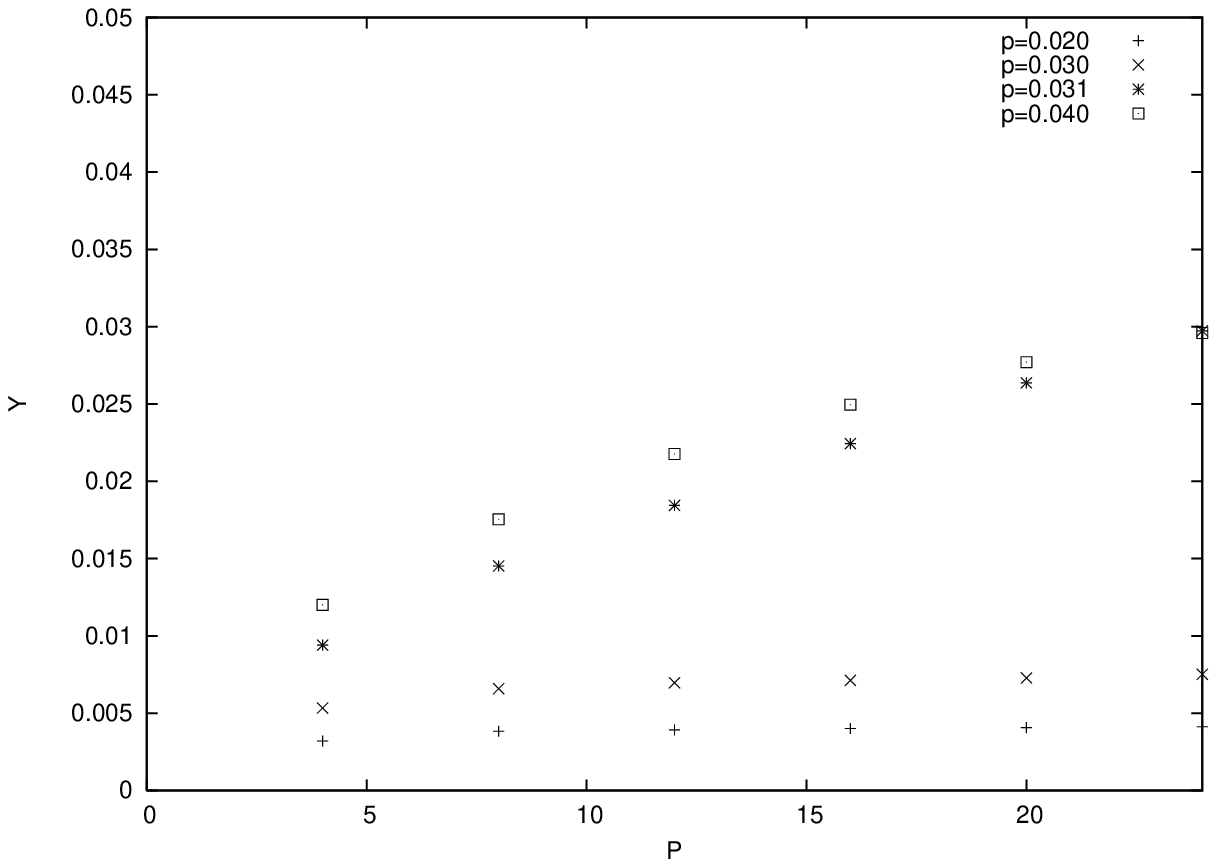}
  \label{fig:14_2lowTbeta=2_0P}
\end{minipage}}
\end{tabular}
\caption{Wilson loop in the low-$T$ region at $\beta=2.0$.
The size of the lattice is $N_v=12^3$.
Simiar plot as in Fig.12. We observe that
 the change in the behavior of $W(C)$ from the area law to 
 the perimeter law takes place at $p\simeq 0.031$.
This critical value almost coincides with $p\simeq 0.033$ 
discussed in Fig.13.
}
\end{figure}

\clearpage


\begin{thebibliography}{99}
\bibitem{kitaev}A.Yu.Kitaev, ``Quantum error correction with imperfect gates",
in {\it Proceedings of the Third International Conference on Quantum
Communication and Measurement}, ed. O.Hirota, A.S. Holevo, and C.M. Caves
(New York, Plenum, 1997);
``Fault-tolerant quantum computation by anyons",
quant-ph/9707021; Annals Phys.303, 2(2003);
M.Freedman, A.Kitaev, M.Larsen, and Z.Wang, ``Topological quantum
computation", quant-ph/0101025.

\bibitem{preskill}J.Preskill, 
``Fault-tolerant quantum computation", quant-ph/9712048; 

\bibitem{dennis}E.Dennis, A.Kitaev, A.Landahl, and J.Preskill,
``Topological quantum memory", quant-ph/0110143;
J.Math.Phys.43, 4452 (2002).

\bibitem{wang}C.Wang, J.Harrington and J.Preskill, 
``Confinement-Higgs transition in a
disordered gauge theory and the accuracy threshold for quantum memory",
quant-ph/0207088; Annals Phys.303, 31 (2003).

\bibitem{mochon}C.Mochon, Phys.Rev.A67, 022315(2003).

\bibitem{benoit}B.Dou\c{o}ot, L.B.Ioffe, and J.Vidal,
``Discrete non-Abelian gauge theories in two-dimensional lattices
and their realizations in Josephson-junction array", cond-mat/0302104.

\bibitem{AI}G.Arakawa and I.Ichinose, ``$Z_N$ gauge theory on a lattice
and quantum memory", quant-ph/0309142; Annals Phys.311, 152(2004).

\bibitem{takeda}K.Takeda and H.Nishimori,
``Self-dual random-plaquette gauge model and the quantum toric code",
hep-th/0310279; Nucl.Phys.B (in press).

\bibitem{kogut}See for example, J.Kogut,
Rev.Mod.Phys.51, 659(1979).

\bibitem{nishimori} See Ref.\cite{wang} and references cited therein;
H.Nishimori, ``{\it Statistical Physics of Spin Glasses and Information
Processing}" (Oxford University Press, 2001).  

\bibitem{estimate}E.Knill, R.Laflamme and W.H.Zurek,
Proc.Roy.Soc.Lond. A 454, 365(1998);
D.Aharonov and M.Ben-Or, 
``Fault-tolerant quantum computation with constant error",
quant-ph/9611025;
``Fault-tolerant quantum computation with constant error rate",
quant-ph/9906129;
A.Yu.Kitaev, Russian Math.Surveys 52, 1191 (1997);
J.Preskill, Proc.Roy.Soc.Lond. A 454, 385(1998);
D.Gottesman, 
``Stabilizer codes and quantum error correction",
quant-ph/9705052.

\bibitem{RBIM}See Ref.\cite{wang} and references cited therein.

\newpage
\end{thebibliography}
\end{document}